# Two-Dimensional Magnetic Resonance Tomographic Microscopy using Ferromagnetic Probes

Mladen Barbic[(a)] and Axel Scherer
*Applied Physics Department and Electrical Engineering Department M/S 200-36*
*California Institute of Technology, Pasadena, California 91125*

We introduce the concept of computerized tomographic microscopy in magnetic resonance imaging using the magnetic fields and field gradients from a ferromagnetic probe. We investigate a configuration where a two-dimensional sample is under the influence of a large static polarizing field, a small perpendicular radio-frequency field, and a magnetic field from a ferromagnetic sphere. We demonstrate that, despite the non-uniform and non-linear nature of the fields from a microscopic magnetic sphere, the concepts of computerized tomography can be applied to obtain proper image reconstruction from the original spectral data by sequentially varying the relative sample-sphere angular orientation. The analysis shows that the recent proposal for atomic resolution magnetic resonance imaging of discrete periodic crystal lattice planes using ferromagnetic probes can also be extended to two-dimensional imaging of non-crystalline samples with resolution ranging from micrometer to Angstrom scales.

(a) mladen@caltech.edu

## INTRODUCTION

Magnetic resonance imaging [1,2] (MRI) has advanced at a rapid pace since initial proposals and demonstrations in 1973 [3,4], with applications in medical imaging attracting the most attention [5,6]. Advances in magnetic resonance microscopy [7] have also been significant [8], recently reaching two-dimensional (2D) imaging resolution of 1μm [9]. Improvements in conventional inductive methods of magnetic resonance detection [10,11] and application of imaging gradients have generally been used to achieve such advances. Although the possibilities of Angstrom-scale resolution were pondered in the early days of MRI [4,12], the ultimate goal of achieving atomic resolution has remained elusive. In 1991, an alternative method of applying the imaging gradients and detecting magnetic resonance, Magnetic Resonance Force Microscopy (MRFM) [13] was proposed, with the ultimate goal of single spin sensitivity and three-dimensional (3D) imaging capability. The technique relies on the atomic scale imaging gradients from the microscopic magnetic particle mounted on a micro-machined mechanical cantilever for the appropriate detection sensitivity required for 3D single spin imaging [14]. Successful MRFM demonstrations were reported for the cases of electron spin [15], nuclear spin [16], and ferromagnetic [17] resonance systems. MRFM research has benefited from the low temperature implementations of the instrument [18], and rapid advances in the fabrication techniques for incorporating smaller magnetic particles [19, 20], and more sensitive mechanical resonators [21]. However, reported MRFM imaging resolution of ~1μm [22,23] remains at the level of inductive detection in conventional MRI.

We recently introduced a complementary atomic resolution magnetic resonance imaging method [24] that significantly relaxes the challenging technical requirements of single spin detection by imaging discrete ordered crystal lattice planes where many spins coherently contribute to the magnetic resonance signal. This approach closely resembles the initial magnetic resonance imaging proposal [4,12] in which linear magnetic field gradients are used to selectively excite magnetic resonance in different atomic lattice planes and produce "diffraction"-like effects. However, our approach differs from this original proposal by introducing non-linear magnetic fields and field gradients from a ferromagnetic sphere to achieve atomic resolution magnetic resonance "diffraction". We investigated various sample-detector coupling mechanisms [25] and showed that the realization of the long-desired atomic resolution magnetic resonance "diffraction" of crystals is well within reach of available experimental techniques. In this article, we present a method that extends the use of imaging gradient magnetic fields from ferromagnetic probes to two-dimensional magnetic resonance microscopy of non-crystalline disordered samples. We show that despite the non-uniform and non-linear nature of the fields from a microscopic magnetic sphere, the concepts of computerized tomography can be applied to obtain proper image reconstruction from the original spectral data by sequentially varying the relative sample-sphere angular orientation. We first review the concept of atomic resolution magnetic resonance "diffraction" of ordered crystals using the fields from a ferromagnetic sphere in order to set the stage for extending the technique to 2-D magnetic resonance tomographic microscopy of non-crystalline samples.

# MAGNETIC RESONANCE IMAGING OF DISCRETE CRYSTAL LATTICE PLANES

Previously described model of atomic resolution magnetic resonance imaging of crystal lattice planes [24,25] relies on placing a ferromagnetic sphere in proximity of the surface of a crystal. A Cobalt sphere, $r_0$=50nm in radius, has a magnetization per unit volume of 1,500 emu/cm$^3$. The simple cubic lattice crystal is assumed to have a unit cell size of $a_0$=3 Angstroms. A large DC magnetic field $B_0$ is applied parallel to the sample surface in the z-direction, polarizing the spins of the atomic lattice as well as saturating the magnetization of the ferromagnetic sphere. A small radio frequency field $B_1$ is applied perpendicular to the large polarizing DC magnetic field $B_0$. The magnetic field from the ferromagnetic sphere at point r in the sample has the following azimuthally symmetric dipolar form:

$$(1) \quad \vec{B}(\vec{r}) = \frac{3\vec{n}(\vec{m}\cdot\vec{n}) - \vec{m}}{|\vec{r}|^3}$$

where n is the unit vector that points from the center of the ferromagnetic sphere to the crystal site location, and m is the magnetic moment vector of the sphere. Since the external DC polarizing magnetic field $B_0$ is considered to be much larger than the field from the ferromagnetic sphere, only the z-component of the magnetic field from the ferromagnetic sphere is included when considering the resonant spins of the atomic lattice [7,24-26]:

$$(2) \quad B_Z(\vec{r}) = \frac{M_0}{|\vec{r}|^3}(3\cos^2\theta - 1)$$

where θ is the angle between the z-axis and the distance vector r, and $M_0$ is the magnitude of the saturation magnetic moment of the ferromagnetic sphere. Figure 1 shows the contours of constant values of the z-component of the magnetic field from the sphere, $B_Z$, that have the azimuthally symmetric form around the z-axis.

In the absence of the ferromagnetic sphere, the discrete nuclear spin sites in the crystal would experience the same externally applied field $B_0$ and therefore meet the magnetic resonance condition at the same magnetic resonance frequency $\omega_R$. However, close to the ferromagnetic sphere, a large magnetic field gradient penetrates into the crystal, and only certain spin sites of the lattice satisfy the correct magnetic resonance conditions at any given magnetic field and frequency values. In the initial model [24,25], a numerical summation was computed to construct a histogram of the number of resonant spin sites in the sample within a 1-Gauss wide shell of constant $B_Z$. This value of the bin width was selected since the linewidth broadening in solids is on the order of 1 Gauss [26]. Distinct spectral peaks were discovered in the number of resonant spin sites with respect to the applied magnetic field in the negative value range, and Figure 2 reproduces the magnetic resonance spectra between the field range of $B_0$–1000 Gauss and $B_0$-500 Gauss for the three cases: (a) semi-infinite crystal, (b) 100-unit-cell thick film, and (c) 100-by-100-by-100 atoms crystallite, with the insets of the figures indicating the sample-sphere relative positions.

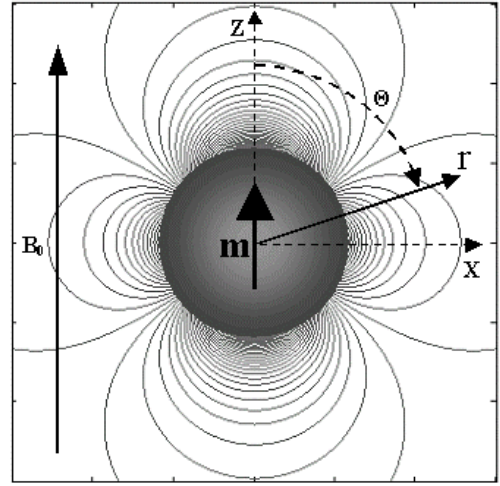

FIG. 1. Azimuthally symmetric contours of constant value of the z-component of the magnetic field $B_Z$ from a ferromagnetic sphere. The spins of the sample laying on the same contour have the same magnetic resonance frequency. For a 100nm diameter Cobalt sphere, the magnetic field gradients are sufficiently large that discrete nature of the spins of the sample needs to be taken into account.

The appearance of spectral peaks was explained using the 3D plots of the resonant spins under the influence of the polarizing magnetic field $B_0$ and the magnetic field from a ferromagnetic sphere [24,25]. Figure 3 shows three such representation plots for crystal lattice spin sites that are in resonance at the magnetic field value of $B_0$–625 Gauss, the location of one of the sharp resonant peaks in the spectra of Figure 2. Only the positive values for the y-axis are plotted for clarity. At this magnetic field value, sections of the shell of constant $B_Z$ perpendicular to the crystal surface intersect the crystal lattice so that a large number of spin sites from the lattice plane at the top and bottom sections of the resonant shell satisfy the resonance condition. The bands of the resonant atoms from the crystal lattice planes are clearly visible, and the resonant ring bands like these at the distinct magnetic field values are responsible for the sharp peaks in the magnetic resonance spectra of Figure 2. It is emphasized that the appearance of the sharp magnetic resonance spectral peaks was only made possible by incorporating into the model the discrete nature of the atomic lattice sites, since magnetic resonance of a continuous medium would result in the monotonic spectrum.

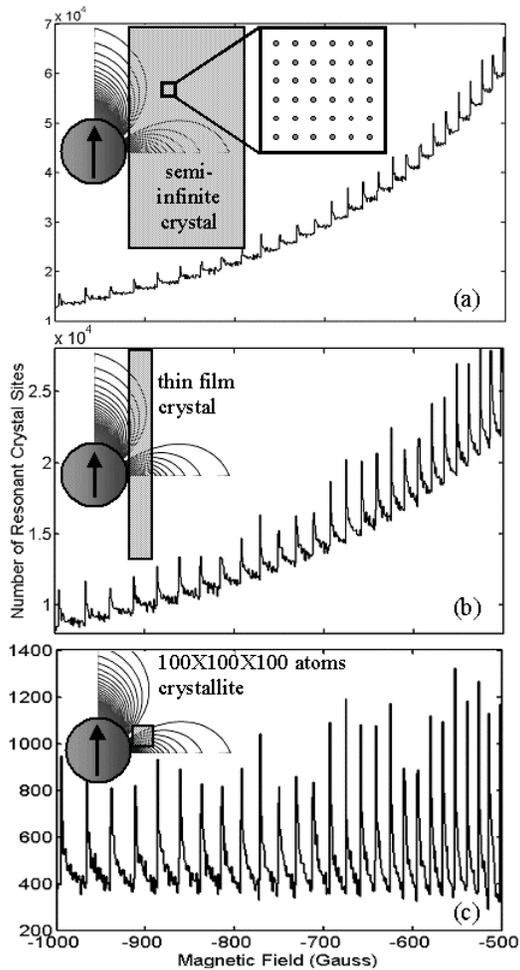

FIG. 2. Magnetic resonance spectra for crystalline samples next to a 100nm diameter Cobalt ferromagnetic sphere for (a) semi-infinite crystal, (b) Thin film 100-unit-cells in thickness, and (c) 100-by-100-by-100 atoms crystallite. Reduction of sample size results in increased spectral peak contrast.

## TOMOGRAPHIC MAGNETIC RESONANCE IMAGING USING FERROMAGNETIC PROBES

Following the description of imaging of crystal lattice planes using ferromagnetic probes, the main question remains as to whether the technique can be extended to the imaging of non-crystalline samples. Inspection of the 3D images of Figure 3, showing the resonant spin sites at a particular value of the magnetic field for crystalline samples with different dimensions, reveals particular features that open the possibility of extending the technique to tomographic imaging of non-crystalline samples. It is apparent from the spectrum of Figure 2(a) and the 3D image of Figure 3(a) that the sharp spectral peaks for the semi-infinite crystal come from the very narrow regions of the sample, while there is a large background signal from other resonant spin sites that are intersected by the 1-Gauss thick shell of constant $B_Z$.

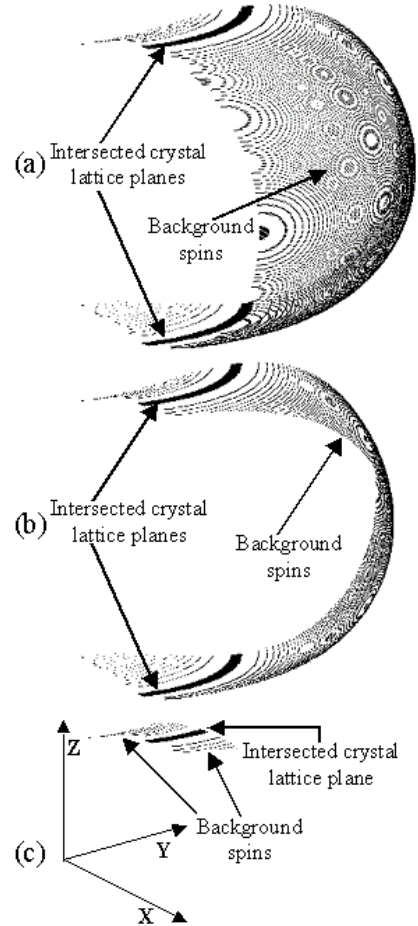

FIG. 3. Three-dimensional plot of the spins in the crystal lattice that are resonant at the magnetic field value of $B_0$-625 Gauss, the location of one of the sharp spectral peaks of Figure 2. Dark bands of spins from the crystal lattice planes perpendicular to the magnetic field direction are responsible for the sharp spectral peaks. For semi-infinite crystal in (a) there is a large background signal that is diminished for the thin crystalline film in (b) since there are no resonant spins beyond the 100th cell in the x-direction. The background signal is further minimized for the case of a small crystallite in (c). Discrete nature of the spins in the crystal lattice is a required for the observation of magnetic resonance spectral peaks.

For the thin film with 100-unit-cell thickness of Figure 2(b) the spectral contrast is significantly increased. The resonant peaks occur at the same location as for the semi-infinite crystal of Figure 2(a), but they lack the large background signal because there are no atoms intersected by the resonant shell beyond the 100th unit cell along the x-direction, as shown in the Figure 3(b). There is still a background signal in Figure 2(b) from the spins that are not part of the crystal lattice planes of interest, but their effect on the spectrum contrast is significantly reduced. The background signal from the spins away from the crystal lattice planes of interest are further minimized for the case of the small crystallite shown in Figures 2(c) and 3(c).

By pursuing this line of argument that the spectral contrast is increased by the reduction of the sample size, we introduce the idea of tomographic magnetic resonance imaging of two-dimensional samples with sizes of ~ 1/10 the size of the ferromagnetic sphere and positioned as shown in Figure 4(a). The sample can represent either a small molecule or protein ~10nm in size underneath a 100nm diameter ferromagnetic sphere, or a biological cell ~10μm in size under a sphere 100μm in diameter. The use of the sphere model and these sphere dimensions is reasonable, as ferromagnetic spheres in this size range have been successfully fabricated [27-30], and have already been integrated as probes on cantilever structures for unrelated applications [31,32]. By magnifying the side view of the model arrangement, as shown in Figure 4(b), it is observed that the sample is intersected by approximately perpendicular lines of constant $B_Z$ along the x-direction. A second view is shown in Figure 4(c) where the front view of the configuration of Figure 4(a) is shown. By taking into consideration the large radius of curvature of the azimuthally symmetric intersecting contours of constant $B_Z$ (shown in Figure 4(a)), it is apparent that the sample is intersected by approximately parallel lines of constant $B_Z$ along the y-direction. Therefore, for an approximately flat two-dimensional sample with lateral dimensions smaller than the ferromagnetic sphere size, the magnetic resonance spectra will be the one-dimensional projections of the spin density along the z-axis. However, as Figure 4(c) shows, the lines of constant $B_Z$ are not equally spaced along the z-axis, since the magnetic field is not linear along the z-axis. Nevertheless, since the magnetic field is well defined in the sample region (equation 2), this non-linearity can be compensated for in the computerized tomographic image reconstruction.

The realization that the magnetic resonance spectrum of the sample in the configuration shown in Figure 4 is a one-dimensional projection of the spin density along the z-axis leads to the possibility of performing a high resolution computerized tomographic reconstruction imaging. Imaging of samples through projections [33] has been an important concept ever since the discovery of x-rays [34], and has been used in the gravitational theory [35] and radio astronomy [36] before becoming widespread through computerized tomography [37] in x-ray [38], electron [39], and optical [40] imaging, among others. Computerized tomographic image reconstruction algorithms are well known [41-43], and here we apply similar principles to the case of magnetic resonance tomographic microscopy using ferromagnetic spheres. By sequentially varying the angle of the sample as shown in Figure 4(c) around the axis shown in Figure 4(b) one can obtain many projections required for computerized tomographic reconstruction.

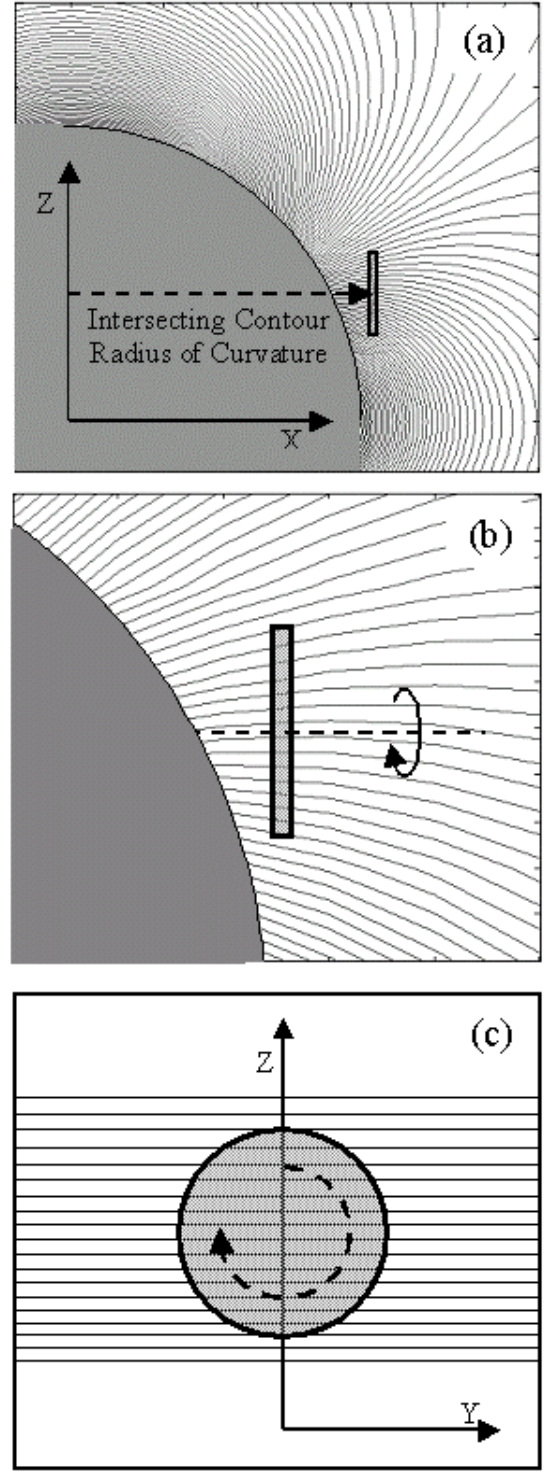

FIG. 4. For the case of a sample small compared to the ferromagnetic sphere and positioned as shown in (a), the contours of constant $B_Z$ are approximately perpendicular to the sample along the x-direction, as magnified in (b). Due to the relatively large radius of curvature of the azimuthally symmetric contours, the sample is intersected by the parallel contours along the y direction as shown in (c). This configuration is suitable for application of computerized tomography methods for image reconstruction.

We note that the image reconstruction from projections in magnetic resonance dates back to the first report of spin distribution imaging [3], and is still performed in the technique of Stray Field Magnetic Resonance Imaging (STRAFI) [44] where constant magnetic field gradients, on the order of 60T/m, from superconducting magnets are used. We demonstrate here that the ultra-high gradient fields from microscopic ferromagnetic probes, as used in Magnetic Resonance Force Microscopy [14] (~5x10$^6$T/m for a 100nm diameter Cobalt sphere), could also be utilized for tomographic imaging of non-crystalline samples with resolution reaching Angstrom levels.

Figure 5 shows the schematic representation of the configuration with the parameters used in the image reconstruction process indicated. In the conventional computerized tomography [39], uniformly separated parallel rays are used to obtain an image projection along an axis, and the one-dimensional Radon transform of the sample density function $\rho(y, z)$ is formed:

$$(3) \quad P_\phi(q) = \int_{(\phi,q)line} \rho(y,z)ds$$

By obtaining a multiple of one-dimensional Radon transforms (3) at different angles ϕ, image reconstruction is performed by the Fourier transform filtered back-projection algorithm for parallel projections [7]:

(4)
$$\rho(y,z) = \int_0^\pi \left\{ \int_{-\infty}^{+\infty} \left[ \int_{-\infty}^{+\infty} P_\phi(q) \cdot e^{i2\pi kq} dq \right] |k| e^{-i2\pi kq} dk \right\} d\phi$$

This reconstruction process involves calculation of the Fourier transform of the Radon transform (innermost bracketed term), multiplication by a ramp function |k| in conjugate space followed by an inverse transformation (outer bracketed term), and finally integration over all angles for the completion of the image reconstruction (outermost integration term) [7].

In this article, a two dimensional phantom is simulated with equally separated 9-by-9 array of spins in the y-z plane. We initially assume that the points are individual proton nuclear spins separated by 5 Angstroms, each having a Lorentzian resonant response with a 10-Gauss linewidth. In the Cartesian coordinate system, the z-component of the magnetic field from a ferromagnetic sphere (equation 2) has the form:

$$(5) \quad B_Z(x,y,z) = \frac{M_0(2z^2 - x^2 - y^2)}{(x^2 + y^2 + z^2)^{5/2}}$$

The center of the phantom array is placed next to the sphere as shown in Figure 4, so that the contours of constant $B_Z$ of equation (5) are approximately perpendicular to the phantom along the x-axis, and parallel to the phantom along the y-axis. This location is determined by the condition that:

$$(6) \quad \frac{\partial B_Z}{\partial x} = \frac{\partial B_Z}{\partial y} = 0$$

and it was previously shown that at this location [25]:

$$(7) \quad x^2 + y^2 = 4z^2$$

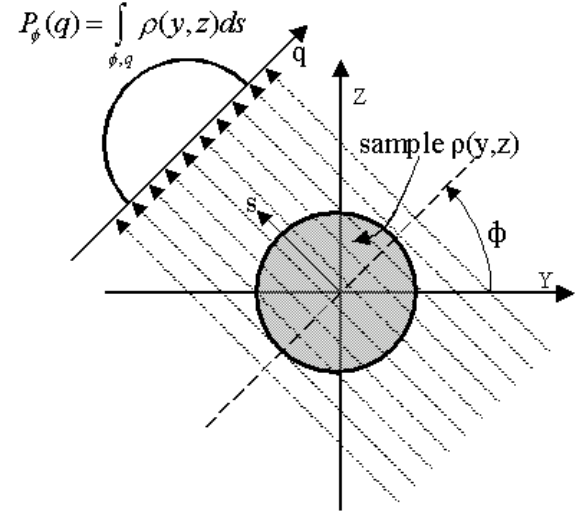

FIG. 5. Configuration and parameters used for conventional computerized tomography. Parallel incident beam is partially absorbed by the sample, and the projection is recorded along the axis q. By recording such projections at many angles Φ, reconstruction of the image can be obtained by using computerized back projection algorithms.

Therefore, when the sample is centered on the y-axis ($y_0$=0), as shown in Figure 4(c), the phantom is positioned along the z-axis as shown in Figure 4(a-b) at the $z_0$ value of:

$$(8) \quad r_0^2 = 4z_0^2$$

Figures 6(b)-(c) show the spectrum of this system of spins under the influence of the large polarizing magnetic field and the field from the ferromagnetic sphere for two imaging angles, 0-degrees and 45-degrees, respectively. There are several observations with respect to Figure 6 that are pertinent to the image reconstruction process. It is immediately apparent that the projection along a single angle is not an ideal projection as described by Equation (3), but is modified by two distorting factors. The first is that the projection in tomographic magnetic resonance microscopy of a single spin is a convolution of the point projection with the Lorentzian lineshape L(q) along the projection line q:

$$(9) \quad R_\phi(q) = \int_{(\phi,q)line} P_\phi(q')L(q-q')dq'$$

The second, and more serious distortion is the non-linearity of the projection due to the non-linear variation of the magnetic field along the z-axis. As already mentioned, in the present configuration, shown in Figure 4, the one-dimensional Radon transforms are not obtained by the uniformly separated parallel rays of constant $B_Z$, but by the contours that have an increasing separation along the z-direction, as Figure 4(c) illustrates.

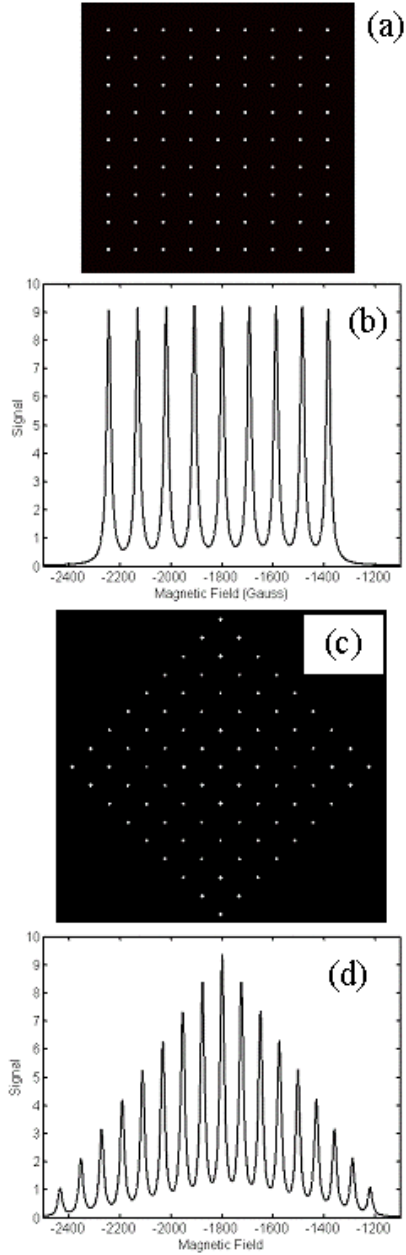

FIG. 6. The simulated phantom consist of a regular array of 9-by-9 spins separated by 5 Angstroms each having a 10-Gauss wide Lorentzian linewidth magnetic resonance response. Under the influence of the large polarizing field and the field from a ferromagnetic sphere, the magnetic resonance spectrum at two respective angles of 0 and 45 degrees are shown. The spectra exhibit a Lorenzian-shape-convoluted projections and distortions due to non-linear variation of the magnetic field from the sphere.

Figure 7(a) shows the projection at 45 degrees superimposed by the ideal projection that one would wish to observe where the response of a spin is a delta-function. As opposed to detecting a series of equally spaced delta functions whose heights represent the number of spins along a projection line, the image is a series of Lorentzian lineshapes that are not evenly spaced and therefore do not represent a true sample projection.

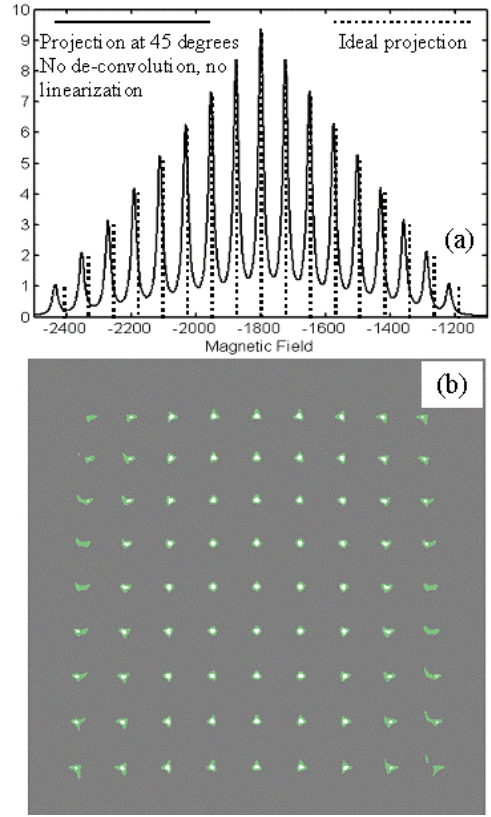

FIG. 7. (a) Projection at 45 degrees compared to the ideal delta-function linear response. (b) Reconstruction from 90 projections at 2 degrees apart without linearization and de-convolution. Non-linear distortions are pronounced at the image edges.

However, despite this distortion, if image reconstruction is performed by the Fourier transform filtered back-projection algorithm for parallel projections of equation (4), one obtains a reasonable reconstruction of the sample spin density, as Figure 7(b) shows. The image was reconstructed from 90 simulated projections at 2 degrees apart. The image is mostly distorted on the edges of the sample where the non-linearity most significantly effects the projection. In addition, since the projection of a single spin is not an ideal delta-function projection, but is a projection convoluted by the Lorentzian lineshape (equation 9), the contrast in the image is also reduced as compared to the original phantom.

Since the magnetic field component (5) used for magnetic resonance imaging in this model is well known, it is simple to correct for the non-linear projection distortion prior to the initial step in the Fourier transform filtered back-projection algorithm. The resulting linearized representative projection at an angle of 45 degrees and the full reconstruction from 90 simulated projections at 2 degrees apart are shown in Figures 8(a) and 8(b), respectively.

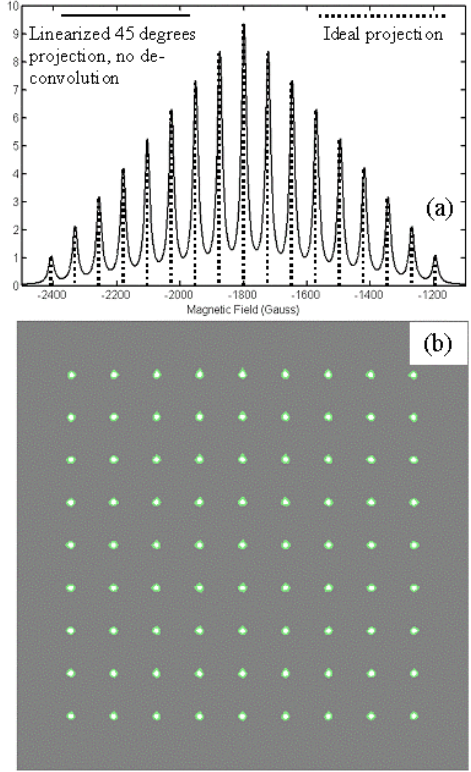

FIG. 8. (a) Linearized projection at 45 degrees compared to the ideal linearized and de-convolved projection. (b) Reconstruction from 90 linearized projections at 2 degrees apart results in a satisfactory reconstructed image with non-linear distortions eliminated.

The computed image is now a very good reconstruction of the original phantom. The reconstructed image still has worse contrast than the original due to the Lorentzian lineshape convolution effect. However, if the single spin response is known, as we assume here for the Lorentzian response for the single spin magnetic resonance, it is still possible to further improve the image reconstruction by: (a) first de-convolving each projection from the Lorentzian lineshape, then (b) perform the linearization, and finally (c) perform the Fourier transform filtered back-projection by using the algorithm for parallel projections of equation (4) to obtain the final image. This de-convolved and linearized projection at a representative angle of 45 degrees is shown in Figure 9(a), and the final computed image from 90 simulated projections at 2 degrees apart is shown in Figure 9(b) where excellent reconstruction of the original image is observed.

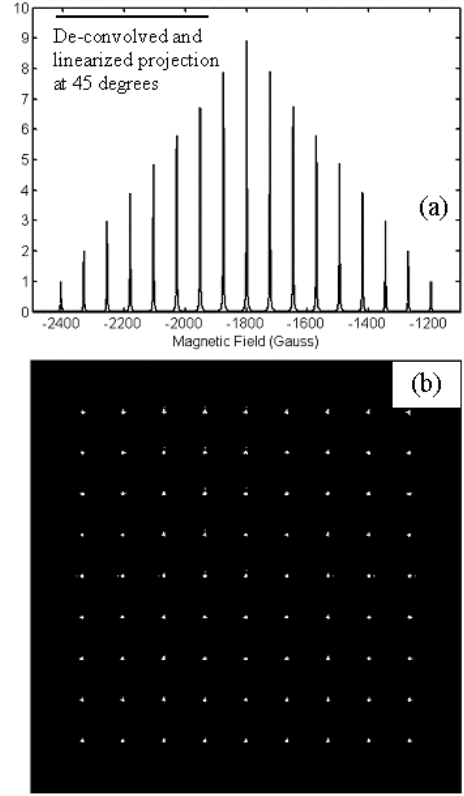

FIG. 9. (a) Linearized and de-convolved projection at 45 degrees, and (b) excellent image reconstruction from 90 projections at 2 degrees apart.

Of course, the intent of this article is to introduce the idea that the two-dimensional tomographic imaging can be performed on samples that are different from the regular ordered arrays of spins. We demonstrate that for the simulated image in Figure 10 where the name of the home institution of the authors is shown. In this case, the projections of the sample will not have regular periodic features as presented so far. Nevertheless, the presented tomographic magnetic resonance imaging method is designed to successfully reconstruct the original image, provided sufficient number of projections is obtained. Figures 10(a)-(c) show the computed images of three successively improved reconstructions, respectively, where 90 simulated projections at 2 degrees apart are used. Figure 10(a) is a direct reconstruction using equation (4) without performing the de-convolution and linearization of projections. As previously described, the image has significant distortions on the edges due to the non-linear distortion of projections. Image 10(b) is a reconstruction of linearized projections that have not been de-convolved from the Lorentzian lineshape, and improved reconstruction is observed. Image 10(c) is a reconstruction that uses both de-convolution and linearization of projections before the final computed reconstruction using equation (4), and excellent reconstruction of the image as well as the contrast are observed.

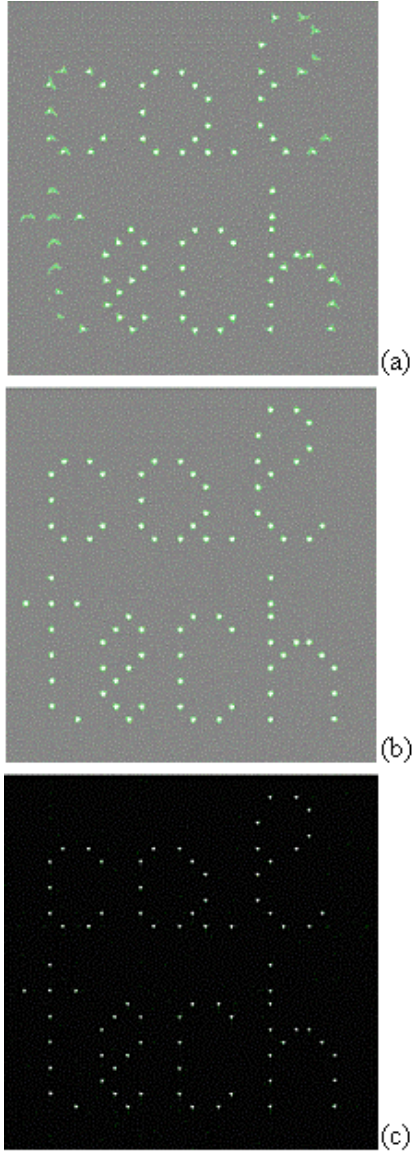

FIG. 10. Image reconstruction of a non-periodic array of spins. In (a) image is reproduced without linearization and de-convolution, and in (b) linearization of projections removes non-linear distortions, while in (c) de-convolution and linearization reproduce the image with excellent contrast.

So far in the article the entire premise has been that the sample is located at the position where the contours of constant $B_Z$ are perpendicular to the sample surface, and positioned with respect to the ferromagnetic sphere as shown in Figure 4. In Figure 11, we show several alternative sample-sphere configurations and their respective sample rotation axis required for computerized tomographic image reconstruction. In all four cases, the contours of constant $B_Z$ are perpendicular to the sample, but vary slightly in the intersecting radius of curvature. Which relative sample-sphere orientation is chosen will depend on the specific experimental and space requirements that might depend on the ferromagnetic probe, detector, RF coils, and polarizing magnet configurations. Detailed analysis of the effect of different configurations on the resolution, sensitivity, linearity, and sample-detector coupling, as well as the potential of extending the technique to three-dimensional computerized tomographic imaging will be presented elsewhere.

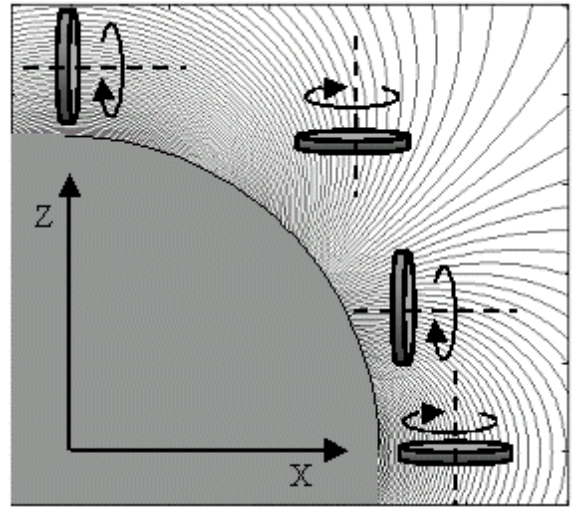

FIG. 11. Several relative ferromagnetic sphere-sample configurations where the contours of constant $B_Z$ are perpendicular to the sample and therefore allow application of computerized tomography techniques for image reconstruction in magnetic resonance microscopy.

## CONCLUSION

We conclude the article by emphasizing that the technique of two-dimensional tomographic magnetic resonance imaging using ferromagnetic probes is distinctly different from other scanning probe techniques that achieve surface atomic resolution imaging, such as scanning tunneling microscopy [45], atomic force microscopy [46], spin polarized tunneling microscopy [47], and spin resonance tunneling microscopy [48,49]. In those methods, the images are obtained by raster scanning an atomically sharp probe over a two dimensional surface of the sample, and at each pixel of the image a measured parameter such as a tunneling current or atomic force is displayed. In the two-dimensional magnetic resonance tomographic reconstruction microscopy, the ferromagnetic probe has no physical atomic scale features to directly achieve atomic resolution, but rather provides the magnetic field gradients that are sufficiently large to atomically resolve magnetic resonance of individual spins along only one direction. The technique does not rely on a point-by-point data acquisition for two-dimensional imaging, but rather is conducted by sequential angular rotation of the sample with respect to the ferromagnetic probe and external field around a prescribed axis. At a single angular orientation and magnetic field value many magnetically resonant spins of the sample are detected simultaneously, but as we demonstrated, by recording such a multi-spin signal at many angular orientations and magnetic field values, it is possible to reconstruct the two-dimensional image representation using the methods of computerized tomography.

**ACKNOWLEDGEMENT**
This work was supported by the Army Research Office grant DAAD 19-00-1-0392, and the Air Force Office for Scientific Research grant F49620-01-1-0497.

**REFERENCES:**
1. P. Mansfield and P. G. Morris *NMR Imaging in Biomedicine* Academic Press, London (1982).
2. C-N Chen and D. I. Hoult *Biomedical Magnetic Resonance Technology* Adam Hilger, Bristol (1989).
3. P. C. Lauterbur, *Nature (London)* **242**, 190 (1973).
4. P. Mansfield and P. K. Grannell, *J. Phys. C: Solid State Phys.* **6**, L422 (1973).
5. M. T. Vlaardingerbroek and J. A. den Boer *Magnetic Resonance Imaging – Theory and Practice*, Springer-Verlag, New York (2003).
6. F. W. Wehrli, *Prog. Nucl. Magn. Reson. Spectrosc.* **28**, 87 (1995).
7. P. T. Callaghan, *Principles of Nuclear Magnetic Resonance Microscopy*, Oxford University Press, New York (1991).
8. J. Aguayo, S. Blackband, J. Schoeniger, M. Mattingly, and M. Hintermann, *Nature (London)* **322**, 190 (1986).
9. S. C. Lee et al., *J. Magn. Reson.* **150**, 207 (2001).
10. D. I. Hoult and R. E. Richards, *J. Magn. Reson.* **24**, 71 (1976).
11. D. I. Hoult and P. C. Lauterbur, *J. Magn. Reson.* **34**, 425 (1979).
12. P. Mansfield and P. K. Grannell, *Phys. Rev. B* **12**, 3618 (1975).
13. J. A. Sidles, *Appl. Phys. Lett.* **58**, 2854 (1991).
14. J. A. Sidles, J. L. Garbini, K. J. Bruland, D. Rugar, O. Zuger, S. Hoen, and C. S. Yannoni, *Rev. Mod. Phys.* **67**, 249 (1995).
15. D. Rugar, C. S. Yannoni, and J. A. Sidles, *Nature (London)* **360**, 563 (1992).
16. D. Rugar, O. Zuger, S. Hoen, C. S. Yannoni, H. –M. Vieth, and R. D. Kendrick, *Science* **264**, 1560 (1994).
17. Z. Zhang, P. C. Hammel, and P. E. Wigen, *Appl. Phys. Lett.* **68**, 2005 (1996).
18. K. Wago, O. Zuger, R. Kendrick, C. S. Yannoni, and D. Rugar *J. Vac. Sci. Technol. B* **14**, 1197 (1996).
19. K. J. Bruland, W. M. Dougherty, J. L. Garbini, J. A. Sidles, and S. H. Chao, *Appl. Phys. Lett*. **73**, 3159 (1998).
20. B. C. Stipe, H. J. Mamin, T. D. Stowe, T. W. Kenny, and D. Rugar, *Phys. Rev. Lett.* **86**, 2874 (2001).
21. T. D. Stowe, K. Yasumura, T. W. Kenny, D. Botkin, K. Wago, and D. Rugar, *Appl. Phys. Lett.* **71**, 288 (1997).
22. O. Zuger and D. Rugar, *Appl. Phys. Lett.* **63**, 2496 (1993).
23. O. Zuger, S. T. Hoen, C. S. Yannoni, and D. Rugar, *J. Appl. Phys.* **79**, 1881 (1996).
24. M. Barbic *J. Appl. Phys.* **91**, 9987 (2002).
25. M. Barbic and A. Scherer *J. Appl. Phys.* **92**, 7345 (2002).
26. C. P. Slichter, *Principles of Magnetic Resonance*, Springer-Verlag, New York (1996).
27. C. Petit, A. Taleb, and M. P. Pileni *Adv. Mater.* **10**, 259 (1998).
28. S. Sun, C. B. Murray, D. Weller, L. Folks, and A. Moser *Science* **287**, 1989 (2000).
29. A. F. Puntes, K. M. Krishnan, and A. P. Alivisatos *Science* **291**, 2115 (2001).
30. T. Hyeon, S. S. Lee, J. Park, Y. Chung, and H. B. Na *J. Am. Chem. Soc.* **123**, 12798 (2001).
31. M. A. Lantz, S. P. Jarvis, and H. Tokumoto *Appl. Phys. Lett.* **78**, 383 (2001).
32. D. R. Baselt, G. U. Lee, K. M. Hansen, L. A. Chrisey, R. J. Colton, *Proc. IEEE* **85**, 672 (1997).
33. S. Webb *From the watching of shadows: the origins of radiological tomography* Adam Hilger, New York (1990).
34. W. C. Röntgen, *Nature (London)* **53**, 274 (1896).
35. J. Radon *Ber. Verh. Sachs. Akad. Wiss.* **69**, 262 (1917).
36. R. N. Bracewell *Austral. J. Phys.* **9,** 198 (1956).
37. A. Cormack *J. Appl. Phys.* **34**, 2722 (1963).
38. G. Hounsfield *British J. Radiol.* **46** 1016 (1973).
39. D. J. de Rosier and A. Klug, *Nature (London)* **217** 130 (1968).
40. D.E. Kuhl and R.Q. Edwards, *Radiol.* **80**, 653 (1963).
41. G. T. Herman *Image Reconstruction from Projections* Academic Press, New York (1980).
42. F. Natterer *The Mathematics of Computerized Tomography* John Wiley & Sons, New York (1986).
43. A. C. Kak and M. Slaney *Principles of Computerized Tomographic Imaging* SIAM, Philadelphia (2001).
44. P. J. McDonald and B Newling *Rep. Prog. Phys*. **61**, 1441 (1998).
45. G. Binning, H. Rohrer, C. Gerber, and E. Weibel, *Phys. Rev. Lett.* **49**, 57 (1982).
46. G. Binning, C. F. Quate, and C. Gerber, *Phys. Rev. Lett.* **56**, 930 (1986).
47. R. Wiesendanger, H. J. Guntherodt, G. Guntherodt, R. J. Gambino, and R. Ruf, *Phys. Rev. Lett.* **65**, 247 (1990).
48. Y. Manassen, R. J. Hamers, J. E. Demuth, and Jr. A. J. Castellano, *Phys. Rev. Lett.* **62**, 2531 (1989).
49. C. Durkan and M. E. Welland, *Appl. Phys. Lett.* **80**, 458 (2002).